\documentclass[prl,aps,twocolumn,showpacs]{revtex4-1}
\usepackage{amsmath}
\usepackage{amssymb}
\usepackage[english]{babel}
\usepackage{graphicx}
\usepackage{bm}
\newcommand{\ket}[1]{\left|#1\right\rangle}

\newcommand{\braket}[3]{\left\langle #1\middle|#2\middle|#3\right\rangle}
\newcommand{\inprod}[2]{\left\langle #1\middle|#2\right\rangle}

\begin{document}

\author{Erik Welander}
\affiliation{Department of Physics, University of Konstanz, 78457 Konstanz, Germany}

\author{Guido Burkard}
\affiliation{Department of Physics, University of Konstanz, 78457 Konstanz, Germany}

\title{Electric Control of the Exciton Fine Structure in Non-Parabolic Quantum Dots}

\begin{abstract}
We show that the non-parabolic confinement potential is responsible for the non-monotonic behavior and sign change of the exciton fine-structure splitting (FSS) in optically active self-assembled quantum
dots.  This insight is important for the theoretical understanding and practical control by electric fields of the quantum state of the emitted light from a biexciton cascade recombination process.  We find that a hard-wall (box) confinement potential leads to a FSS that is in better agreement with experimentally measured FSS than a harmonic potential.  We then show that a finite applied electric field can be used to remove the FSS entirely, thus allowing for the creation of maximally entangled photons, being vital to the growing field of quantum communication and quantum key distribution. 
\end{abstract}
%
%
%
%
\maketitle

\textit{Introduction}.
Entangled photons, a non-classical state of light, are an indispendible part of proposed and implemented protocols for optical quantum communication and quantum key distribution \cite{gisin}. Parametric down-conversion (PDC) is a well-studied and established way of creating entangled photons but suffers from two major drawbacks: since on the order of $10^{10}$ pump photons are required per created photon pair, the process is rather inefficient. Secondly, the creation time is highly stochastic. Most protocols, however, require an on-demand source of fixed photon number, leading to the search for alternatives to PDC. One of the most promising candidates is the biexciton cascade recombination from a single semiconductor quantum dot (QD) \cite{benson} considered in this paper.

\begin{figure}[t]
\includegraphics[scale=0.7]{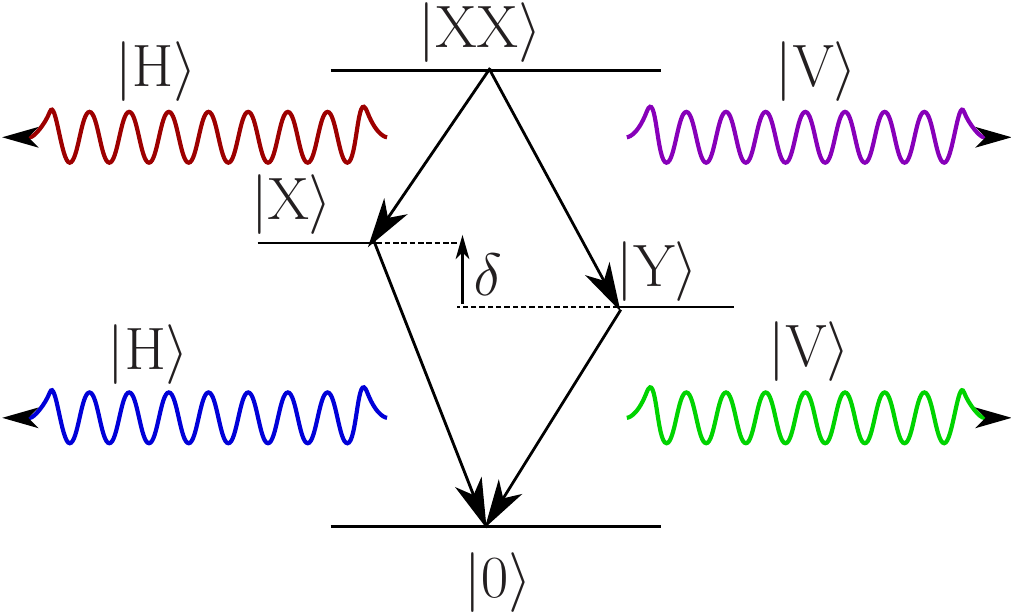}
\caption{\label{decay}The bixeciton recombination cascade. For degenerate intermediate exicton levels ($\delta \ll \Gamma$, where $\Gamma$ is the linewidth of the emitted light and $\delta$ the FSS) the two decay paths are indistinguishable and the recombination can be described  by $\ket{\mbox{XX}}\ket{0}\rightarrow(\ket{\mbox{X}}\ket{\mbox{H}}+\ket{\mbox{Y}}\ket{\mbox{V}})/\sqrt{2}\rightarrow\ket{\mbox{0}}(\ket{\mbox{HH}}+\ket{\mbox{VV}})/\sqrt{2}$ where we use the notation $\ket{\mbox{electronic}}\ket{\mbox{photonic}}$ and $\ket{XX}$ is the biexciton state, $\ket{X}$ ($\ket{Y}$) denotes the exciton state with horizontal (vertical) polarization, $\ket{H}$ ($\ket{V}$) the photonic state with horizontal (vertical) polarization and $\ket{0}$ is the electronic ground state or photon vacuum.}
\end{figure}

The biexciton is the bound state of two electrons and two holes in a semiconductor. Upon radiative recombination, one electron and one hole are annihilated by the emission of a single photon, leaving a single exciton. As a next step in the cascade, the exciton recombines and emits a second photon. Because of the Pauli exclusion principle the ground state of the biexciton is a singlet, whereas there are two possbile exciton states, characterized by the underlying $p_x$ or $p_y$ orbitals of the hole. Thus the emitted light can have either horizontal, $\ket{H}$, or vertical, $\ket{V}$, polarization when emitted from the biexciton and from the intermediate exciton. If the two exciton states are energetically indistinguishable, the two recombination paths are equivivalent with respect to the frequency of the emitted light, and the two photons are entangled in polarization, being described by the state $(\ket{HH} + \ket{VV})/\sqrt{2}$. The process is schematically shown in Fig.~\ref{decay}.

For real QDs, however, the intermediate exciton states are typically not degenerate, but split by an energy $\delta$ known as the exciton fine-structure splitting (FSS). Since the two recombination paths now have energetically different intermediate levels, a simple frequency measurement could reveal the ``which-way'' information, and thus the entanglement is, at least partially, lost \cite{santori}. Understanding of the FSS is essential for any model considering the entanglement of the light produced by the biexciton cascade recombination and there are several different sources contributing to the FSS, which have been the topic of a number of experimental and theoretical studies in recent years. These include intrinsic effects caused by the underlying crystal orientation \cite{bester,karlsson,stock,singh}, piezoelectric effects \cite{bimberg} and the QD geometry \cite{bayer,hawrylak,tong}. There is a fundamental physical difference between the intrinsic splitting and geometry dependent splitting. As pointed out by Singh and Bester \cite{singh}, for the intrinsic case, the FSS decreases for smaller dots, as opposed to the geometry dependent effect, where a smaller dot gives rise to a larger FSS. We claim that these two statements do not stand in conflict with each other, since they describe two different mechanisms. The intrinsic splitting is caused by the actual shape of the underlying atomic orbitals whereas the geometry dependent one is mediated by the electron-hole overlap. This means that the intrinsic FSS depends on the number of atomic orbitals, which increases for larger dots, in contrast to the geometry dependent FSS which scales with the electron-hole overlap that decreases with increasing dot size.  
In this paper, we presume that the crystal orientation has already been chosen to eliminate the intrinsic FSS and focus on the geometry dependent splitting caused by the long-range electron-hole exchange interaction and determined by the in-plane shape of the QD.   We especially investigate the possibility of restoring degeneracy by virtue of applying an in-plane electric field. The dependence of the FSS on an in-plane electric field has been experimentally studied and it was confirmed that the FSS can be removed \cite{gerardot,vogel} but a complete theoretical understanding is still missing. Earlier work using a harmonic dot confinement potential find qualitatively different results \cite{hawrylak}. Here we use a more realistic confinement potential and compare the obtained FSS to the one from harmonic confinement and to experimental results.  Our theory explains the observed FSS in InGaAs QDs including the sign change of the FSS when a lateral electric field was applied \cite{vogel}.  
In addition to the application of lateral electric fields, there are other methods of reducing the FSS by strain \cite{bester}, vertical electric
fields \cite{bennett,bennett2,ghali}  or a magnetic field \cite{stevenson}, as well as proposals to recover the entanglement in the presence of a finite FSS, such as spectral filtering \cite{akopian,hohenester}, time reordering \cite{avron} or embedding the QD in an optical cavity \cite{stace}.

\textit{Theoretical model}.
We consider a quantum dot composed of one cubic semiconductor surrounded by a material composed of another cubic semiconductor. The electronic structure is characterized by bands of which we consider the conduction band states, labelled $m_s = \pm 1/2$, and the valence band states, labelled $m_j = \pm 3/2$ for heavy holes and $m_j = \pm 1/2$ for light holes. Because of the large distance in energy to these valence band states we do not consider the split-off band. The surrounding material is required to have a conduction(valence) band of higher(lower) energy at the $\Gamma$-point than the dot material. We start from a general Hamiltonian, describing two particles in a semiconductor,
\begin{equation}
\label{origeq}
 H_0 = H_1(\mathbf{r}_1) + H_1(\mathbf{r}_2) + V(\mathbf{r}_1 - \mathbf{r}_2),
\end{equation}
where $H_1(\mathbf{r})$ is the single particle Hamiltonian,
\begin{equation}
H_1(\mathbf{r}) =  \frac{p^2}{2m} + V_{\textrm{lattice}}(\mathbf{r}),
\end{equation}
with the underlying periodic lattic potential, $V_{\textrm{lattice}}(\mathbf{r})$, and $V(\mathbf{r}_1 - \mathbf{r}_2) = e^2/4\pi\epsilon_0\epsilon_r|\mathbf{r}_1 - \mathbf{r}_2|$, the Coulomb potential with the relative dielectric constant $\epsilon_r$. In the $\mathbf k \cdot \mathbf p$ method a basis for the electron wave functions is formed from the Bloch waves as
\begin{equation}
\varphi_{s\mathbf k}(\mathbf r) = e^{i\mathbf k \cdot \mathbf r}u_s(\mathbf r),
\end{equation}
where $s$ is a band index and $u_{s}(\mathbf r) = u_{s, \mathbf k = \mathbf 0}(\mathbf r)$ is the Bloch function at the $\Gamma$-point of band $s$. For the study of two-particle systems an anti-symmetric wave function is needed and thus we form the two-particle basis
\begin{equation}
\varphi_{s\mathbf k_1t\mathbf k_2}(\mathbf r_1, \mathbf r_2) = \frac{\varphi_{s\mathbf k_1}(\mathbf r_1)\varphi_{t\mathbf k_2}(\mathbf r_2) - \varphi_{s\mathbf k_1}(\mathbf r_2)\varphi_{t\mathbf k_2}(\mathbf r_1)}{\sqrt{2}}.
\end{equation}
Inserting this into Eq.~(\ref{origeq}), we find a $\mathbf k$ dependent Hamiltonian containing an interband $\mathbf k \cdot \mathbf p$ term.  Treated as a perturbation up to second order, these terms alone give rise to the effective mass approximation. As shown by Pikus and Bir \cite{pikus}, including the Coulomb exchange between particles leads to more correction terms such as the band diagonal first order Hartree correction,
\begin{equation}
\label{hartree}
H^C_{s^\prime\mathbf k_1^\prime t^\prime\mathbf k_2^\prime,s\mathbf k_1t\mathbf k_2} = \delta_{ss^\prime}\delta_{tt^\prime}V_{\mathbf k_1^\prime-\mathbf k_1}\delta_{\mathbf k_1 + \mathbf k_2,\mathbf k_1^\prime + \mathbf k_2^\prime},
\end{equation} with $V_{\mathbf k}$ the Fourier transform of the Coulomb  potential as well as the third order exchange term, $H^a$, with elements
\begin{align}
\label{exchange}
\begin{split}
\braket{s^\prime\mathbf k_1^\prime t^\prime \mathbf k_2^\prime}{H^a}{t\mathbf k_2s\mathbf k_1} 
= &\sum_{\alpha\beta}\frac{\hbar^2}{m^2}\frac{p^\alpha_{s^\prime t}\,p^\beta_{t^\prime s}}{E_g^2}\left(k_1^\alpha - {k_2^\prime}^\alpha\right)\times\\
\times \left(k_1^\beta - {k_2^\prime}^\beta\right) 
& V_{\mathbf k_1^\prime - \mathbf k_2}\delta_{\mathbf k_1 + \mathbf k_2,\mathbf k_1^\prime + \mathbf k_2^\prime},
\end{split}
\end{align}
where $E_g$ is the band gap energy, $\alpha$ and $\beta$ run over the spatial coordinates, and $\inprod{\mathbf r_1,\mathbf r_2}{t\mathbf k_2s\mathbf k_1} = \varphi_{s\mathbf k_1}(\mathbf r_2)\varphi_{t\mathbf k_2}(\mathbf r_1)$. 
With the exchange interaction known for basis vectors, the two-electron wave functions can be expressed as
\begin{align}
\begin{split}
\Psi_{st}(\mathbf r_1,\mathbf r_2) = & \sum_{\mathbf k_1, \mathbf k_2} c_{s\mathbf k_1, t\mathbf k_2}\varphi_{s\mathbf k_1t\mathbf k_2}(\mathbf r_1, \mathbf r_2) \\ 
= &\frac{1}{\sqrt 2}\sum_{\mathbf k_1, \mathbf k_2} c_{s\mathbf k_1, t\mathbf k_2}e^{i(\mathbf k_1\cdot\mathbf r_1 + \mathbf k_2\cdot\mathbf r_2)}u_s(\mathbf r_1)u_t(\mathbf r_2) \\
- & \frac{1}{\sqrt 2}\sum_{\mathbf k_1, \mathbf k_2} c_{s\mathbf k_1, t\mathbf k_2}e^{i(\mathbf k_1\cdot\mathbf r_2 + \mathbf k_2\cdot\mathbf r_1)}u_s(\mathbf r_2)u_t(\mathbf r_1) \\
= & \frac{\psi_{st}(\mathbf r_1,\mathbf r_2) - \psi_{st}(\mathbf r_2,\mathbf r_1)}{\sqrt 2} = \inprod{\mathbf r_1, \mathbf r_2}{st}.
\end{split}
\end{align}
We choose
\begin{equation}
\psi_{st}(\mathbf r_1, \mathbf r_2) = \psi_s(\mathbf r_1)\psi_t(\mathbf r_2),
\end{equation}
where
\begin{equation}
\psi_s(\mathbf r) = \sum_{\mathbf k}c_{s\mathbf k}e^{\mathbf k \cdot \mathbf r}u_s(\mathbf{r}) = F_s(\mathbf r)u_s(\mathbf{r}),
\end{equation}
and $F_{s}(\mathbf r)$ is the envelope function which is the solution without corrections (\ref{hartree}) and (\ref{exchange}) to the equation
\begin{equation}
\label{enveq}
-\frac{\hbar^2\nabla^2}{2m_s^\ast}F_{s}(\mathbf r) + V_s(\mathbf r)F_{s}(\mathbf r) = \epsilon_sF_{s}(\mathbf r),
\end{equation}
where $m_s^\ast$ is the effective mass of band $s$ and $V_s(\mathbf r)$ is a band dependent potential varying on a mesoscopic scale describing the confining structure such as quantum dot. The new correction terms, (\ref{hartree}) and (\ref{exchange}) can now be expressed in terms of envelope functions and are equal to
\begin{align}
\label{coulombintegral}
\begin{split}
&\braket{s^\prime t^\prime}{H^C}{st} \\
= &\delta_{ss^\prime}\delta_{tt^\prime}\int\left|F_s(\mathbf{r_1})\right|^2\left|F_t(\mathbf{r_2})\right|^2V(\mathbf r_1 - \mathbf r_2)\,d\mathbf r_1\,d\mathbf r_2
\end{split}
\end{align}
and
\begin{align}
\label{exchangeintegral}
\begin{split}
\braket{s^\prime t^\prime}{H^a}{st}  = & -\sum_{\alpha\beta}\frac{\hbar^2}{m^2}\frac{p^\alpha_{s^\prime t}p^\beta_{t^\prime s}}{E_g^2}
\int \frac{\partial^2 V(\mathbf r_1 - \mathbf r_2)}{\partial r^\alpha_1 \partial r^\beta_1}\times\\
\times & F^\dagger_{s^\prime}(\mathbf r_1)F_t(\mathbf r_1)F^\dagger_{s}(\mathbf r_2)F_{t^\prime}(\mathbf r_2) d\mathbf r_1\,d\mathbf r_2.
\end{split}
\end{align}
A similar expression was presented by Kadantsev and Hawrylak \cite{hawrylak} as well as by Tong and Wu \cite{tong}. If we let $t^\prime,t$ represent valence band electrons we need to reverse the order and apply the time-reversal operator when going over from the electron-electron picture to the electron-hole picture. This amounts to
\begin{equation}
H^{a(e-h)}_{s^\prime t^\prime,st} = -H^{a(e-e)}_{s^\prime\Theta t,s\Theta t^\prime},
\end{equation}
where $\Theta$ is the time-reversal operator \cite{pikus}.

A suitable potential, $V_s(\mathbf r)$, has to be choosen to properly describe the system under consideration, in our case the quantum dot. We discuss different choices in the next section. For a given $V_s(\mathbf r)$, we numerically solve Eq.~(\ref{enveq}) to find the envelope functions. This provides us with the single particle states $\ket{s = m_s}$ and $\ket{t = m_j}$ for electrons and holes from which exciton product states are formed as $\ket{m_s, m_j} = \ket{m_s} \otimes \ket{m_j}$. Using these envelope functions, the integrals Eqs.~(\ref{coulombintegral}) and (\ref{exchangeintegral}) are calculated numerically and the following exciton eigenvalue problem is formulated in the basis of $\ket{m_s, m_j}$ as
\begin{equation}
\label{excitonequation}
\left(H^0 + H^C + H^a\right)\ket{X} = E_X\ket{X},
\end{equation}
with $H^0$ being the excitation and confinement energies
\begin{equation}
H^0_{m_s^\prime m_j^\prime,m_sm_j} = \delta_{m_sm_s^\prime}\delta_{m_jm_j^\prime}(E_g + \epsilon_{m_s} + \epsilon_{m_j}),
\end{equation}
in which $\epsilon_s,\; s \in \{m_s, m_j\}$ are the energies from Eq.~(\ref{enveq}). This eigenvalue problem is solved numerically and two of the eigenvectors, $\ket{X_x}$ and $\ket{X_y}$ are identified as the ones having maximal projections on $\ket{\sigma_+} + \ket{\sigma_-}$ and $\ket{\sigma_+} - \ket{\sigma_-}$ respectively, where $\ket{\sigma_\pm} = \ket{\mp 1/2, \pm 3/2}$ for heavy excitons and $\ket{\sigma_\pm} = \ket{\pm 1/2, \pm 1/2}$ for light excitons. The fine structure splitting is now calculated as
\begin{equation}
\delta = E_x - E_y,
\end{equation}
with exciton energies $E_x,E_y$ taken from Eq.~(\ref{excitonequation}).

\textit{Results}.
We now discuss two choices of confinement potentials, $V_s(\mathbf r)$.
The objective is to describe the confinement of electrons and holes 
to a given nanostructure. In our case we are interested in a quantum dot with dimensions $l_x \times l_y \times l_z$. Typically we let $l_x \neq l_y$ which causes a FSS. Physically the confinement comes from a change in the underlying composition such as going from InGaAs to GaAs. It is necessary to take a band-dependent potential into account since electrons and holes are usually subject to different band offsets. Further, we also include an electric field which is represented by $-q_s\mathbf E \cdot \mathbf r$ in the Hamiltonian, where $q_s$ is the charge of a particle in band $s$, i.e. $-e(+e)$ for conduction (valence) bands. 
Here, we restrict ourselves to heavy, bright excitons, i.e. $\ket{\mp 1/2, \pm 3/2}$.

\textit{Harmonic potential}.
A simple model for the QD is the harmonic confining potential 
\begin{equation}
V_s(\mathbf r) = \left(\sum_{\alpha}\frac{m^\ast_s\omega_{s\alpha}^2}{2}r^2_\alpha\right) 
-q_s\mathbf E \cdot \mathbf r. 
\end{equation}
The solutions are harmonic oscillator wave functions with characteristic lengths
$l_{s\alpha} = \sqrt{\hbar/m_s^\ast\omega_{s\alpha}}$ defining the spread of the wave function and $l^{\mathbf E}_{s\alpha} = q_sE_\alpha/m_s^\ast\omega_{s\alpha}^2$, the electric displacement. 
\begin{figure}[t]
\includegraphics{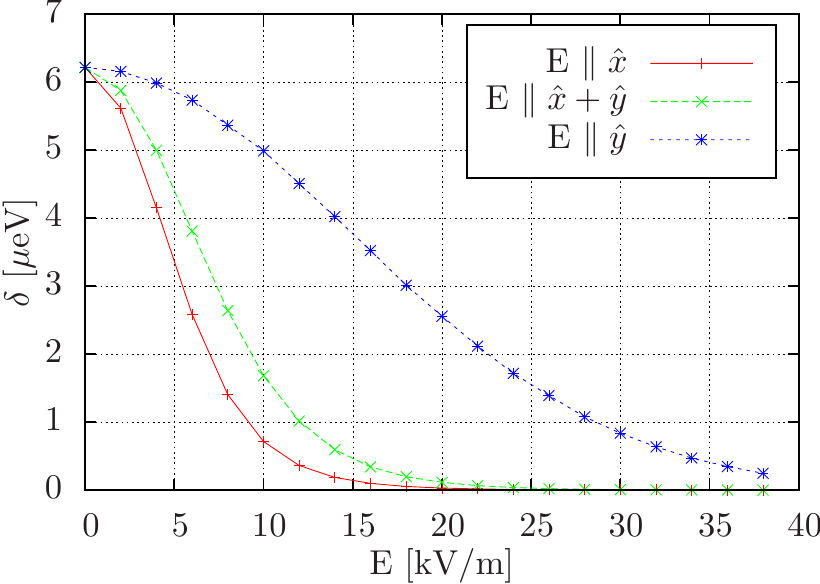}
\caption{\label{figharmonicFSS}The FSS $\delta$
calculated for an harmonically confined InGaAs dot with charecteristic lengths $30 \times 20 \times 7$ nm$^3$ and various directions of the electric field. Regardless of the direction of the field, the FSS is monotonically decreasing and vanishes only asymptotically.}
\end{figure}
The FSS in this case as a function of the electric field is plotted in Fig.~\ref{figharmonicFSS}. We observe that the FSS is decreasing with increasing field, but does not change sign.  Experiments, however, show another picture including a non-monotonic behaviour as well as a change of sign \cite{vogel}.   This is because in a hamonically confined dot, the electron and hole can be arbirarily separated only due to the electric field. This neglects actual ``hard wall'' confinement of the dot, which is not affected by the electric field, and this suggests another model is necessary to properly understand the experimentally observed FSS.

\begin{figure}[t]
\includegraphics[scale=0.92]{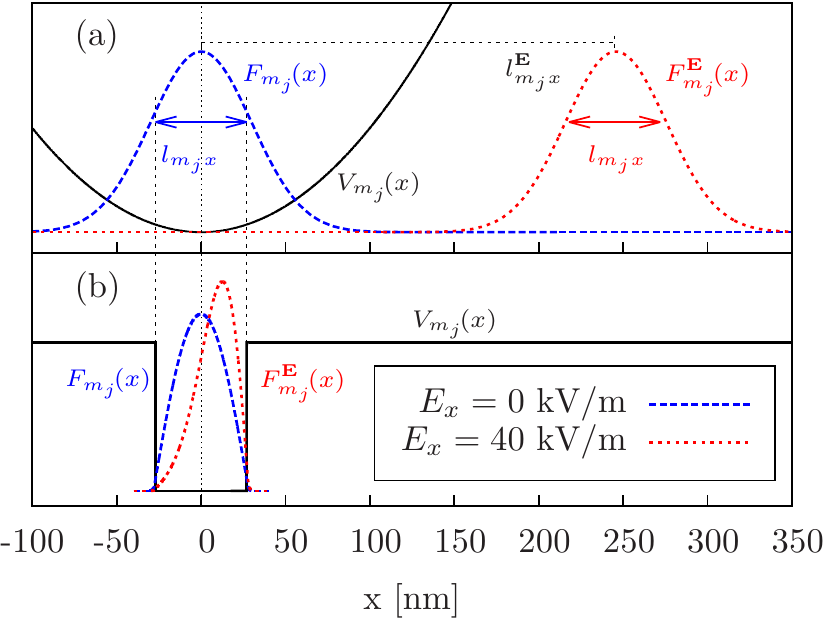}
\caption{\label{figwaves}Hole envelope functions with, $F^\mathbf{E}_{m_j}(\mathbf r)$, and without, $F_{m_j}(\mathbf r)$, applied electric field $E_x$ in x-direction, for (a) harmonic and (b) hard-wall confinement.  For the case of harmonic confinement, the wave functions are translated by $\l^{\mathbf E}_{m_j}$ and are no longer inside the physical region of the dot when a field $E_x$ is applied but retain their shape. For the hard-wall potential, the main effect of the electric field is to deform the wave functions; this is accompanied by a relatively small shift that leaves the particles inside the dot.   The deformation affects the second derivative of   $F^\mathbf{E}_{m_j}(\mathbf r)$ which determines the FSS, see Eq.~(\ref{exchangeintegral}).}
\end{figure}

\textit{Hard-wall confinement}.
To incorporate the effects of a physical confinement we a consider rectangular box of dimensions $l_x \times l_y \times l_z$ which has a potential step at the boundary, i.e.
\begin{equation}
V_{s}(\mathbf r) = -\Delta E_s\left(\prod_\alpha\chi_{[-l_\alpha/2,l_\alpha/2]}(r_\alpha)\right)
-q_s\mathbf E \cdot \mathbf r,  
\end{equation}
where $\Delta E_s$ is the band offset and $\chi_{A}(r)$ is the characteristic function of the set $A$.
\begin{figure}[t]
\includegraphics[scale=0.95]{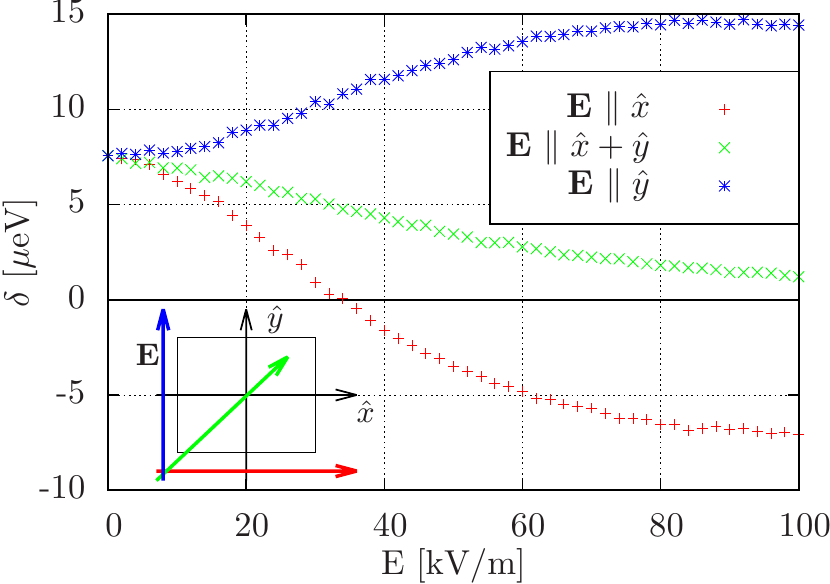}
\caption{\label{figvogel}The FSS $\delta$  
calculated for In$_{0.2}$Ga$_{0.8}$As/GaAs dot of dimensions $55 \times 50 \times 7$ nm$^3$ with a hard-wall confinement in the presence of an electric field applied in various directions. The FSS is sensitive to the direction in which the field is applied: In the $\hat x$-direction the FSS decreases and changes sign at a critical field strength where $\delta = 0$ (here, at $E_x\simeq 35\,{\rm kV/m}$). When the field is applied in the $\hat y$-direction, the FSS \emph{increases} instead.}
\end{figure}
As can be seen in Fig.~\ref{figvogel} the FSS now exhibits more structure and depends more drastically on the direction of the applied field.  A comparison between the envelope functions found for the two different cases reveals that the harmonically confined particles are displaced but not deformed by the electric field, whereas for the case of a hard-wall confinement, the shape of the envelope function is modified as well. Fig.~\ref{figwaves} shows the effect for the hole envelope function. When an electric field is applied, the harmonically confined particles move outside physical boundaries of the dot, whereas the particles confined by a step potential remain inside the box. By inspecting Eq.~(\ref{exchangeintegral}) we also note that exchange integral depends not only on the electron-hole overlap but also on the curvature of the envelope wave function.  The curvature, in turn, depends on $E$ in the case of the hard-wall confinement, but not for the harmonic confinement. 

\textit{Conclusions}.
We theoretically investigated the exciton FSS for quantum dots of cubic semiconductor materials and its dependence on dot geometry and applied electric field. We found that the choice of confinement potential in the model is of great importance and we noted that different confinement potentials can lead to qualitatively different results. A model with a harmonic confinement potential cannot capture the experimentally observed features including a sign change of the FSS under the application of an electric field.   Using a more realistic potential step (hard wall) barrier, we find a more complex relation between FSS and field and predict the possibility of a complete suppression of the FSS, as observed experiments \cite{vogel}. We trace the additional FSS structure back to the fact that electron and hole wave functions are not only displaced but also deformed by the hard-wall potential.  This deformation influences the FSS via the second derivative of the envelope function.  The suppression of the FSS by means of an electric field allows the creation of entangled photons without additional post processing which is of interest to the field of quantum communication. Open questions include the effects of an applied vertical field as well as the influence of light holes.

We gratefully acknowledge fruitful discussions with Bill Coish as well as funding by the Konstanz Center of Applied Photonics (CAP), SFB 767, and BMBF QuaHL-Rep.

\end{document}